\documentclass[aps,pra,showpacs,twocolumn,groupedaddress]{revtex4}

\usepackage{graphicx,bbm,amsthm, amsmath, amssymb}
\usepackage{color}

\def\tr{\,{\rm tr}\,}
\def\ket#1{|#1\rangle}
\def\bra#1{\langle#1|}

\def\bracket#1#2#3{\langle #1 |#2|#3 \rangle}
\def\ie{{i.e.}}
\def\eg{{e.g.}}
\def\norm#1{{\parallel\! #1 \!\parallel}}
\def\ii{{\rm i}}

\def\etal#1{#1}
\def\tit#1{}

\begin{document}

\title{Initial-state randomness as a universal source of decoherence}

\author{Marko \v Znidari\v c}

\affiliation{
Department of Physics, Faculty of Mathematics and Physics, 
University of Ljubljana, 
SI-1000 Ljubljana, Slovenia}

\date{\today}

\begin{abstract}
We study time evolution of entanglement between two qubits, which are part of a larger system, after starting from a random initial product state. We show that, due to randomness in the initial product state, entanglement is present only between directly coupled qubits and only for short times. Time dependence of the entanglement appears essentially independent of the specific hamiltonian used for time evolution and is well reproduced by a parameter-free two-body random matrix model.
\end{abstract}


\maketitle

\section{Introduction}

In the last few years quantum entanglement is one of the most active research fields in quantum physics, for a review see Refs.~\cite{entReview}. On microscopic level of coherent quantum systems entanglement can be used in various quantum protocols to perform non-classical operations. On the other hand, in the world of macroscopic objects with many degrees of freedom at high temperatures there is apparently no observable manifestation of entanglement. There have been many attempts to explain this classicality of macroscopic systems. Most notably, decoherence due to external degrees of freedom is usually credited as being responsible for the disappearance of entanglement from macroscopic superpositions. In a nutshell, the argument goes as follows: even if the system is in an entangled state at the beginning, \eg, in a coherent superposition of two macroscopic states, time evolution will in general transform this coherent superposition into an incoherent (\ie, classical) mixture. The reason for such decoherence is an always present residual coupling of our central system to many uncontrollable external degrees of freedom -- the environment. For a review of decoherence see Ref.~\cite{decoherence}. However, one must be aware that the evolution of the central system plus environment is still unitary and therefore, even though the final state of the central system and environment will presumably be very complex, in principle, it will be a pure state possessing some bipartite entanglement. 

The resolution of this apparent paradox is similar to the one with the second law of thermodynamics~\cite{Lebowitz}. Increasing of the thermodynamic entropy with time is seemingly in contradiction with the reversibility of the underlying equations of motion. For explanation one can use two observations: (i) practicality -- performing time reversal by, \eg, reversing velocities of all particles might be close to impossible from a practical point of view; (ii) probability -- initial conditions are prevalently of such form that in almost all cases the entropy will subsequently increase. Similar arguments can be used to explain the apparent lack of entanglement in quantum systems with many degrees of freedom. First, even though joint pure state representing central system and environment is bipartite entangled, the detection of entanglement might be close to impossible because it would require very complex measurements involving very many degrees of freedom. Indeed, using entanglement witnesses it has been shown that the detection of entanglement in a sufficiently complex state gets exponentially hard with increasing number of particles~\cite{ranW}. For all practical purposes the detection of entanglement in such states is impossible. Second argument, that is the role played by initial conditions in the course of loosing entanglement by time evolution is the subject of present work.

We are going to study how the entanglement between two qubits changes during hamiltonian time evolution. Hamiltonian evolution will act on a system of totally $n$ qubits, two of which will be chosen as our central system of interest while the remaining $n-2$ will act as the ``environment''. The idea is to study how a general hamiltonian evolution changes entanglement of a smaller subsystem, whose degrees of freedom we presumably are able to control and therefore also measure its entanglement. Time evolution with a general hamiltonian, say quantum chaotic one, will in general produce states whose statistical properties are well described by those of random quantum states. For random quantum state on $n$ qubits one knows~\cite{iden} that tracing out $n-2$ qubits will, for large $n$, with high probability result in a separable two qubit reduced density matrix. Therefore, sufficiently ``complex'' time evolution will eventually wipe out entanglement between two qubits. How are things then, for instance, in integrable systems, which in general do not generate completely random states? One point we have not touched so far is the role played by initial conditions. For integrable systems there can exist simple initial states for which entanglement will persist also for long times, nevertheless, as we will see, the majority of initial conditions is such that entanglement between two qubits will rapidly decay with time {\em irrespective of the hamiltonian}. This universality will be a consequence of the generic form of initial states -- their randomness.  

The initial pure state will be chosen to be a product state on the central system (two qubits) and either random or random product state on the rest. Therefore, initially there will be no entanglement between the two chosen qubits. We are then going to study how much entanglement can be produced by various hamiltonian evolutions and how long will it take until it disappears. Entanglement will change with time due to two competing effects. One is entanglement production due to time evolution with non-separable hamiltonian, while the other is entanglement loss due to the spreading of initial state randomness throughout the system and the approach of system's state to a random state. The net result will be the increase of entanglement at short times and a complete lack thereof after some critical time. In addition, time dependence of entanglement will turn out to be universal, that is independent of the specifics of the hamiltonian used in time evolution.

There have been many studies of entanglement evolutions, let us here mention only those that are closer to the present work and deal with two qubit entanglement. Evolution of concurrence for initial product states has been studied for XY model in magnetic field in Ref.~\cite{Monta:03}, see also Ref.~\cite{Rossignoli:07}. For sufficiently strong coupling between qubits concurrence initially increased with time after which it rapidly decreased to zero, similarly as in the present work. Evolution of entanglement for initially entangled states of two qubits (Bell states) which are weakly coupled to a generic environment has been studied in~\cite{Pineda:07}. In such cases entanglement monotonically decreases with time from its maximal value at time zero. Evolution of entanglement for initial Bell state in an integrable XY model has been studied in~\cite{Amico:04}. Our hamiltonian will be homogeneous in space and therefore the two qubits of the central system will be coupled. This must be contrasted to studies of the so-called environment induced entanglement generation~\cite{envinduced}, where two qubits are uncoupled. Disappearance of entanglement (on average) after finite critical time found in the present work is reminiscent of the so-called sudden death of entanglement, where initially entangled state of two uncoupled qubits becomes separable after a finite time of open system dynamics~\cite{ESD}. A system consisting of two spins coupled to electrons has been studied for initial separable states in Ref.~\cite{Gao:05}, while time evolution of entanglement for initial thermal states in a XY model has been considered in Refs.~\cite{Sen:05,Huang:06}. For concurrence in a kicked Ising model see~\cite{Arul}. In~\cite{Sen:06} it has been found that starting from an initial non-disordered product state in a spin glass model entanglement can persist for long times. Von Neumann entropy of a block of spins in Ising model and for product initial state has been studied in~\cite{Dur:05}.

\section{Quantifying entanglement}    

We are going to study entanglement between two qubits which are in turn part of a larger $n$ qubit system. For two qubits positivity of partially transposed density matrix with respect to one qubit, $\rho^{T_A}$, is a necessary and sufficient condition for its separability~\cite{PPT}. Negative eigenvalues (for two qubit states there can be at most one) therefore signal the presence of entanglement. A quantity measuring this is negativity~\cite{neg} $N(\rho)$ which is equal to the sum of absolute values of negative eigenvalues of $\rho^{T_A}$ and can be defined as
\begin{equation}
N(\rho)=\frac{\norm{\rho^{T_A}}_1-1}{2},
\label{eq:negdef}
\end{equation}
with the trace norm $\norm{A}_1=\tr{\sqrt{A^\dagger A}}$. For two qubit density matrices it is simply $N(\rho)=|\lambda^{T_A}_{\rm min}|$ if the minimal eigenvalue $\lambda^{T_A}_{\rm min}$ is negative and $0$ otherwise.   

Entanglement of formation~\cite{Fully}, which quantifies quantum resources needed to create a given state, has nicer mathematical properties as negativity (or logarithmic negativity). For two qubit systems entanglement of formation $E_{\rm F}(\rho)$ can be calculated in terms of a simpler quantity called concurrence $C(\rho)$, defined as
\begin{equation}
C(\rho)={\rm max}\,\{ 0, \lambda_1-\lambda_2-\lambda_3-\lambda_4\},
\label{eq:Cdef}
\end{equation}
where $\lambda_i$ are square roots of decreasingly ordered eigenvalues of $(\rho\, \sigma^y\otimes \sigma^y \rho^* \sigma^y \otimes \sigma^y)$, calculated in standard computation basis. $E_{\rm F}(\rho)$ is then given by\cite{Cdef}
\begin{equation}
E_{\rm F}(\rho)=H\left(\frac{1+\sqrt{1-C^2(\rho)}}{2}\right),
\label{eq:Efdef}
\end{equation}
with $H(x)=-x \log_2{x}-(1-x)\log_2{(1-x)}$ being a binary entropy. For pure states the entanglement of formation is given by the von Neumann entropy of the reduced density matrix, while it is defined by a convex roof extension (minimization over all convex realizations of $\rho$) for mixed states. A state is separable iff its concurrence or iff its negativity is zero. 

The third and last quantity used in measuring entanglement will be the so-called fully entangled fraction~\cite{Fully} defined as
\begin{equation}
f(\rho)={\rm max}\, \bracket{\psi}{\rho}{\psi},
\label{eq:fdef}
\end{equation}
where maximization runs over all maximally entangled states obtained by local unitary transformations from maximally entangled state, \ie,  $\ket{\psi}=U_1 \otimes U_2 (\ket{00}+\ket{11})/\sqrt{2}$. One can distill maximally entangled singlets from an ensemble of $\rho$ using BBPSSW~\cite{BBPSSW} protocol iff $f>1/2$. Fully entangled fraction $f$ can be used as a lower bound for the entanglement of formation~\cite{Fully},
\begin{equation}
E_{\rm F} \ge h(f),
\end{equation}
where $h(f)$ is expressed in terms of binary entropy $H(x)$,
\begin{displaymath}
h(f)= \left\{ 
\begin{array}{ll}
H(\frac{1}{2}+\sqrt{f(1-f)}) & , f \ge \frac{1}{2} \\
0 & , f < \frac{1}{2}
\end{array}
\right. .
\end{displaymath}
In the above inequality an equal sign holds if $\rho$ is pure state. Note that $f\le 1/2$ does not necessarily mean that the entanglement of formation is zero. Fully entangled fraction also determines maximal teleportation fidelity~\cite{maxTeleport}. Fully entangled fraction $f(\rho)$ is equal to the largest eigenvalue of the real part of the density matrix $\rho$ written in the Bell basis, in which all maximally entangled states have real expansion coefficients. If $f \ge 1/2$ this is in turn equal to~\cite{Horodecki_fully}
\begin{equation}
f(\rho)=\frac{1}{4}(1+\Theta(\rho)),\qquad \Theta(\rho)=\norm{T}_1=\tr{\sqrt{T^\dagger T}},
\label{eq:taudef}
\end{equation} 
where $T$ is a real $3\times 3$ dimensional correlation matrix given by $T_{ij}=\tr{(\rho\, \sigma^i \otimes \sigma^j)}$, with $\sigma^i$ being Pauli matrices, \ie, $i,j \in \{x,y,z\}$. Because of its simple analytical form we are going to study $\Theta(\rho)$ (\ref{eq:taudef}) rather than $f(\rho)$ (\ref{eq:fdef}). They essentially give the same information in the interesting regime of $f>1/2$. If $\Theta(\rho) >1 $ then the state $\rho$ can be used in entanglement distillation. Norm of the correlation matrix $T$ can be used as a simple entanglement criterion also for many-qubit systems~\cite{Badziag:08}. 

In the following we are therefore going to study negativity $N(\rho)$ (sometimes just minimal eigenvalue of the partially transposed matrix $\lambda^{T_A}_{\rm min}$), concurrence $C(\rho)$ and $\Theta(\rho)$ which is connected to the fully entangled fraction. As we will see, qualitatively all behave in the same way. From the analytical viewpoint though $\Theta(\rho)$ is the simplest quantity and is therefore the best candidate for an analytical treatment.

\section{Systems studied}

We are going to study entanglement evolution for various one dimensional spin hamiltonians consisting of $n$ spin-1/2 particles. To verify that the results do not depend on the underlying dynamics we will use both chaotic and integrable systems.

Heisenberg spin-$1/2$ model is an integrable model with hamiltonian
\begin{equation}
H=\sum_i \sigma_i^x \sigma_{i+1}^x+\sigma_i^y \sigma_{i+1}^y+\sigma_i^z \sigma_{i+1}^z.
\label{eq:heis}
\end{equation}
We have checked that the results are similar for anisotropic Heisenberg model as well as for isotropic Heisenberg model in a tilted magnetic field.

We can break integrability of the Heisenberg model by applying magnetic field, for instance, a staggered field in $z$-direction,
\begin{equation}
H=\sum_i \sigma_i^x \sigma_{i+1}^x+\sigma_i^y \sigma_{i+1}^y+\sigma_i^z \sigma_{i+1}^z+\sum_i h_i \sigma_i^z,
\label{eq:heisS}
\end{equation}
where the strength of the magnetic field is $h_{2i}=0$ and $h_{2i+1}=-\frac{1}{2}$ on odd sites. 
\begin{figure}[h]
\centerline{\includegraphics[width=3.3in]{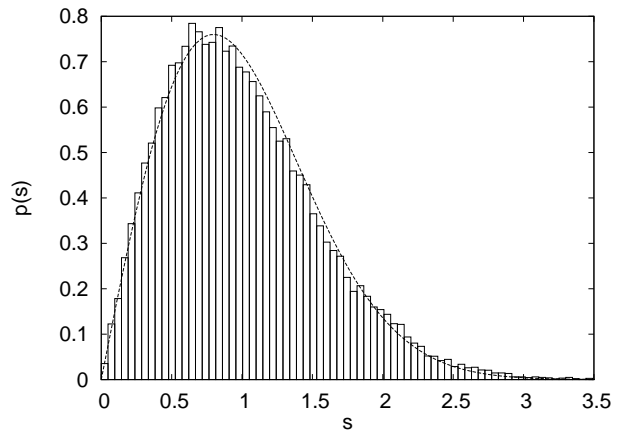}}
\caption{Level spacing distribution for the Heisenberg model in a staggered field (\ref{eq:heisS}) for $n=16$. Dashed line is Wigner-Dyson distribution holding for quantum chaotic systems.}
\label{fig:lsdHeisstag}
\end{figure}
As we can see in Fig.~\ref{fig:lsdHeisstag}, spacing of neighboring energy levels agrees with the Wigner-Dyson distribution, $p(s)=s \pi/2 \exp{(-s^2 \pi/4)}$, which approximates distribution of spacings for gaussian orthogonal random matrix ensemble and is typical for quantum chaotic systems~\cite{rmt}. 

Last model will be Ising chain in tilted magnetic field, 
\begin{equation}
H=\sum_i \sigma_i^x \sigma_{i+1}^x+\sigma_i^x+\sigma_i^z.
\label{eq:TI}
\end{equation}
Ising model in tilted magnetic field also displays typical signatures of quantum chaos~\cite{TIsing}, similarly as its time-dependent kicked version~\cite{Prosen:00}. We have checked that the results are essentially the same also for integrable transverse Ising model.

As we will see, all three models will display similar evolution of entanglement being in turn equal to the one for a two-body random matrix model. Therefore, our main focus will actually be on a two-body random matrix model, for which we have only nearest neighbor coupling terms,
\begin{equation}
H=\sum_i h_{i,i+1},
\label{eq:rmt}
\end{equation}  
with $h_{i,i+1}$ acting nontrivially only on two qubits, for which it is a $4 \times 4$ random hermitian matrix, same for all coupled pairs and normalized as $\tr{(h_{i,i+1}^2)}=1$. A random hermitian matrix is a matrix whose matrix elements are independent random complex gaussian numbers~\cite{rmt}. We always averaged over an ensemble of random matrices $h_{i,i+1}$. 

For all hamiltonians the geometry is that of an one-dimensional chain with open boundary conditions. The state at time $t$ is obtained as $\ket{\psi(t)}=\exp{(-\ii H t)} \ket{\psi(0)}$ from which we get the reduced density matrix for the two qubits between which we study entanglement,
\begin{equation}
\rho(t)=\tr_{n-2}{\ket{\psi(t)}\bra{\psi(t)}},
\label{eq:rho}
\end{equation}
where a subscript $n-2$ means tracing over $n-2$ qubits. The above $\rho(t)$ will then be used in calculating various entanglement measures. Two qubits in question will be either nearest neighbors, that is qubits directly coupled by the hamiltonian, or two qubits which are not directly coupled, \eg, next nearest neighbors. Typically they will be chosen in the middle of the chain with the results being largely independent of their precise location. The initial pure state will be of two forms. Most of the time we are going to consider random product initial state,
\begin{equation}
\ket{\psi(0)}=\ket{\chi}_1 \otimes \cdots \otimes \ket{\chi}_n,
\label{eq:prodic}
\end{equation}
where $\ket{\chi}_i$ is a random state of $i$th qubit, given by
\begin{equation}
\ket{\chi}_i=\cos{\phi_i} \, {\rm e}^{\ii \alpha_i}\ket{0}_i+\sin{\phi_i}\, {\rm e}^{\ii \beta_i}\ket{1}_i,
\label{eq:random1}
\end{equation}
with $\beta_i,\alpha_i,\phi_i$ independent random numbers given as $\alpha_i=2\pi \xi$, $\beta_i=2\pi \xi$ and $\phi_i=\arcsin{\sqrt{\xi}}$ where all three $\xi$s are independent (for each qubit) uniform random numbers on interval $[0,1]$.

\section{Entanglement evolution}

\subsection{Perturbative expansion}
Initially, at time $t=0$, our initial state is always of product form between two qubits in question and therefore there won't be any entanglement, \ie, $C(t=0)=0$, $\lambda^{T_A}_{\rm min}(t=0)=0$ and $\Theta(t=0)=1$. Subsequent evolution will entangle two qubits therefore one expects that the entanglement will gradually build up. For sufficiently short times, one can use perturbation theory to calculate the reduced density matrix $\rho(t)$ (\ref{eq:rho}). Taking for $H$ nearest-neighbor hamiltonian with an arbitrary two-qubit coupling term $h^{(2)}$ we get after expanding propagator $\exp{(-i H t)}$ to the lowest order in time,
\begin{equation}
\lambda^{T_A}_{\rm min}=-|\delta| t\,+{\cal O}(t^2),\qquad \delta=\bracket{\chi_A^\perp \chi_B^\perp}{h^{(2)}}{\chi_A \chi_B}.
\label{eq:lminLR}
\end{equation}
In the derivation of the above formula we assumed the initial product state on the two qubits in question while an arbitrary state is allowed on the remaining $n-2$ qubits, $\ket{\psi(0)}=\ket{\chi_A}\otimes\ket{\chi_B}\otimes\ket{\chi}_{n-2}$. States $\ket{\chi^\perp_{A,B}}$ are single qubit states orthogonal to $\ket{\chi_{A,B}}$ and because only absolute value of $\delta$ enters their phases do not matter. Similar calculation for concurrence (\ref{eq:Cdef}) and $\Theta$ (\ref{eq:taudef}) gives,
\begin{eqnarray}
C&=&2|\delta|t\,+{\cal O}(t^2) \nonumber \\
\Theta&=&1+4|\delta|t\,+ {\cal O}(t^2). 
\label{eq:CLR}
\end{eqnarray}
We can see that the initial speed at which entanglement is produced depends only on a single matrix element of $h^{(2)}$ between the initial product state $\ket{\chi_A}\otimes \ket{\chi_B}$ and the corresponding orthogonal product state. Because for all quantities the initial time scale depends trivially on the value of $|\delta|$ we will measure time in rescaled dimensionless units
\begin{equation}
\tau=t \overline{|\delta|},
\label{eq:tau}
\end{equation}
with $\overline{|\delta|}$ being the average absolute value of a matrix element, where averaging is done over random initial single qubit states $\ket{\chi_{A,B}}$ (\ref{eq:random1}). For the isotropic Heisenberg model (\ref{eq:heis}) we get $\overline{|\delta|}=1$, for the Heisenberg model in a staggered field (\ref{eq:heisS}) one gets $\overline{|\delta|}\approx 0.8882$ and for the Ising model in a tilted magnetic field (\ref{eq:TI}) $\overline{|\delta|}\approx 0.6168$. For two-body random matrix model (\ref{eq:rmt}) we can, instead of averaging over initial product states, average over ensemble of random matrices $h^{(2)}$, resulting in $\overline{|\delta|}=\sqrt{\pi}/4\approx 0.4431$. In all figures showing time dependence of entanglement we are going to use dimensionless time $\tau$.

\subsection{Numerical results}

\begin{figure}[h]
\centerline{\includegraphics[width=3.3in]{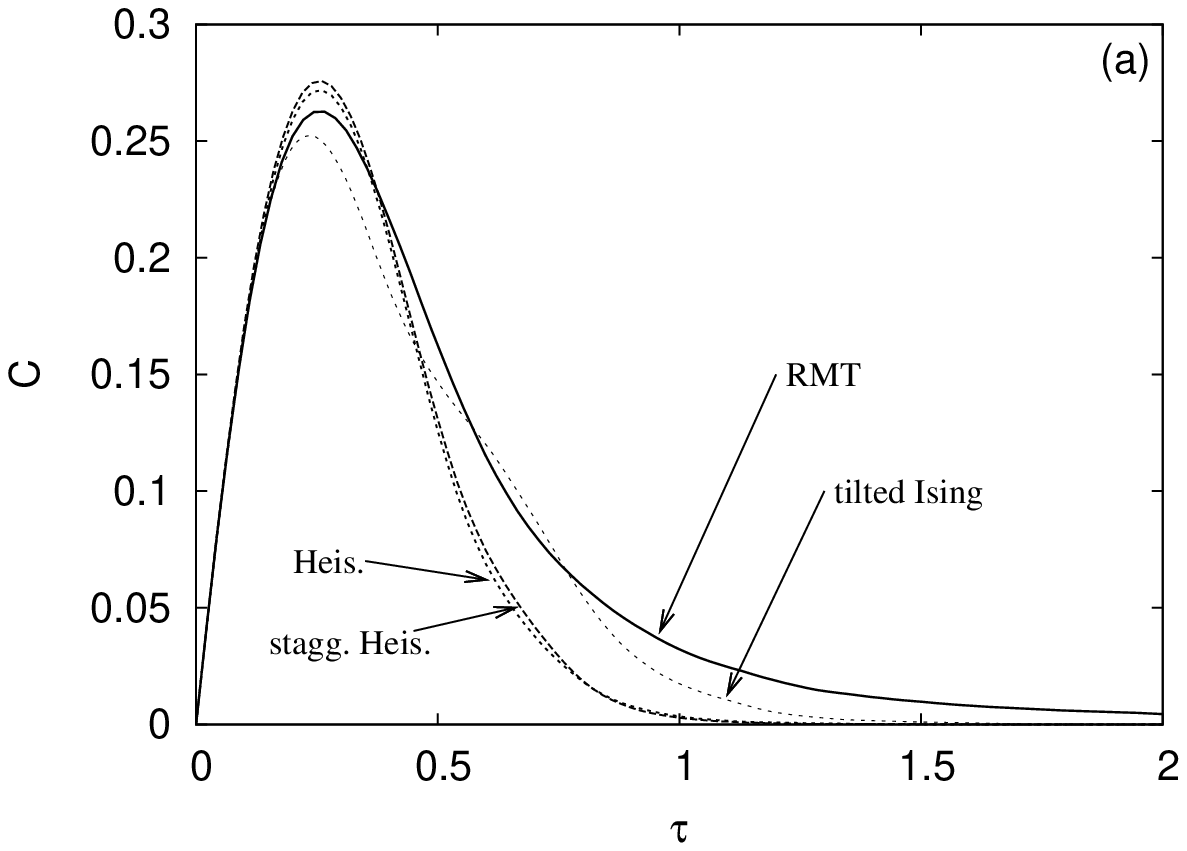}}
\centerline{\includegraphics[width=3.3in]{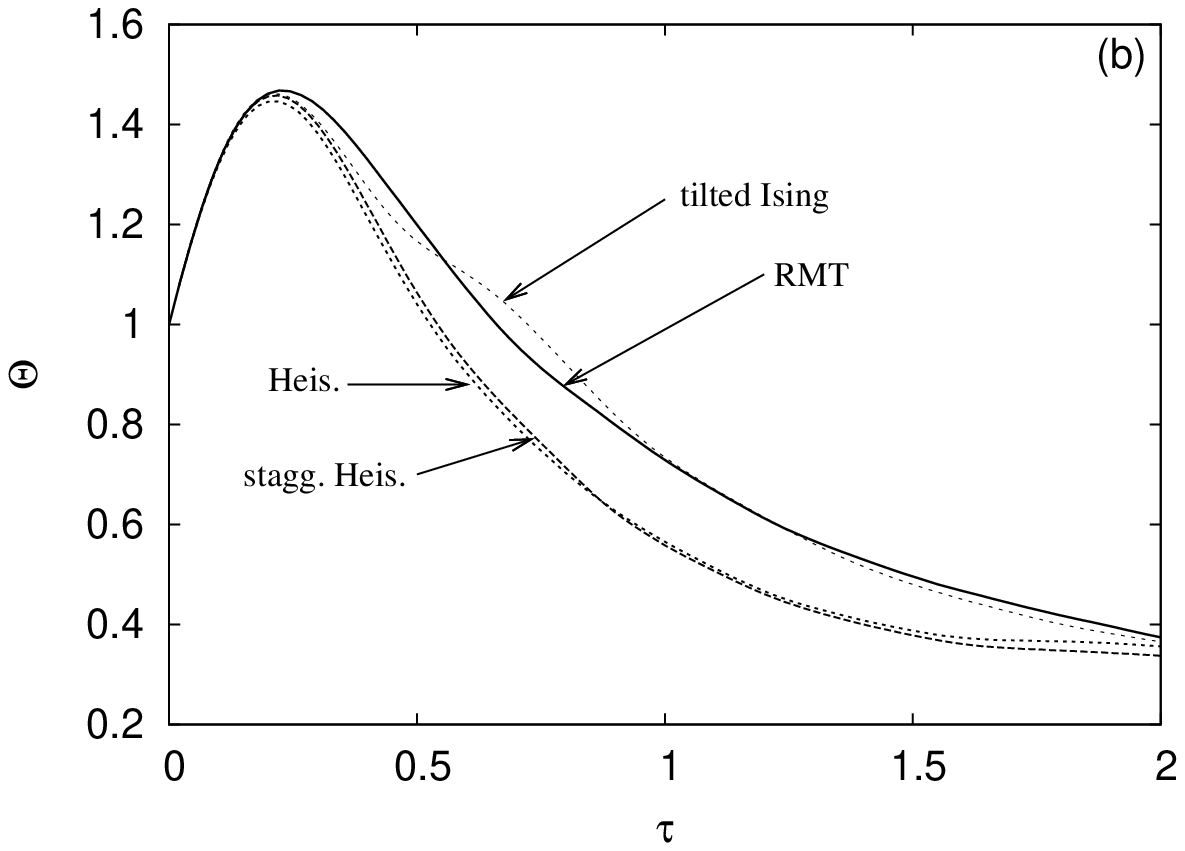}}
\caption{Average concurrence $C(\tau)$ (\ref{eq:Cdef}) in Fig.(a) and average $\Theta(\tau)$ (\ref{eq:taudef}) in Fig.(b) for various hamiltonians. Thick full curves are for two-body random matrix model (\ref{eq:rmt}), thin dotted curves for the Ising model in a tilted magnetic field (\ref{eq:TI}), short dashed curves are for isotropic Heisenberg model (\ref{eq:heis}) while long dashed curves are for the Heisenberg model in a staggered field (\ref{eq:heisS}). Averaging is performed over random product initial states (\ref{eq:prodic}) for $n=16$. Times $\tau^*$ when $\Theta=1$ are $0.66$ for two-body random matrix model, $0.73$ for tilted Ising and $0.53$ for both Heisenberg models.}
\label{fig:H}
\end{figure}

First, we performed numerical simulations of time evolution, calculating average $C(\tau)$, $\lambda_{\rm min}^{T_A}(\tau)$ and $\Theta(\tau)$ for different hamiltonians. Averaging has been done over random initial product states (\ref{eq:prodic}). As one can see in Fig.~\ref{fig:H} the behavior is overall similar for all studied hamiltonians. For instance, from Fig.~\ref{fig:H}b we can see that for times larger than some critical $\tau^*$ the average $\Theta$ is below $1$ which means that the two qubits can not be used for distillation any more. Critical $\tau^*$ is for all models between $0.5$ and $0.75$. Similar dependence (not shown) is also obtained for the average minimal eigenvalues of partially transposed density matrix $\lambda_{\rm min}^{T_A}$, which also becomes positive at roughly the same $\tau^*$. Concurrence, seen in Fig.~\ref{fig:H}a, has a similar time dependence. The only difference is that for $\tau > \tau^*$ concurrence is not strictly zero but instead decays exponentially with time. This is a consequence of the fact that even though for $\tau> \tau^*$ the two qubits are on average not entangled any more there are still exponentially rare instances (product initial states) for which there is still some entanglement present. With time the probability of such entangled states decreases exponentially. In all cases dependence for small times agrees with analytical perturbative result for concurrence in eq.~\ref{eq:CLR} and Fig.~\ref{fig:H}a, for $\Theta$ in eq.~\ref{eq:CLR} and Fig.~\ref{fig:H}b, and for $\lambda_{\rm min}^{T_A}$ in eq.~\ref{eq:lminLR} and Fig.~\ref{fig:lmin}.

Note that $\tau^*$ is a time when the average quantity (like $\Theta$ or $\lambda_{\rm min}^{\rm T_A}$) reaches a certain value ($1$ or $0$). It should not be confused with the average time $\bar{\tau}^*$ when $\Theta$ (or $\lambda_{\rm min}^{\rm T_A}$ or $C$) reaches $1$. For each individual initial condition time when a state gets separable, {\em i.e.} critical $\tau$, is of course the same as that of $C$ or $\lambda_{\rm min}^{\rm T_A}$. However, becouse distributions change with time, time $\tau^*$ is not exactly the same as the average time $\bar{\tau}^*$. The average times $\bar{\tau}^*$ are for $\lambda_{\rm min}^{\rm T_A}$ (as well as for $C$ or negativity) equal to $\bar{\tau}^*=0.72$ for two-body random matrix model and tilted Ising model and $\bar{\tau}^*=0.59$ for both Heisenberg models. The average times $\bar{\tau}^*$ for $\Theta(\tau)$ are on the other hand slightly different, $\bar{\tau}^*=0.64$ for two-body random matrix model, $\bar{\tau}^*=0.61$ for tilted Ising model and $\bar{\tau}^*=0.47$ for both Heisenberg models. Compare these $\bar{\tau}^*$ with $\tau^*$ listed in Figs.~\ref{fig:H}b,~\ref{fig:f} and~\ref{fig:lmin}. One can observe that $\bar{\tau}^*$ for $\Theta$ and $\lambda_{\rm min}^{\rm T_A}$ are slightly different, for instance in the case of a two-body random matrix model $0.64$ vs. $0.72$. The fact that $\bar{\tau}^*$ is for $\Theta$ smaller than for $\lambda_{\rm min}^{\rm T_A}$ is not a contradiction as $f \le 1/2$ does not necessarily mean that the entangelement of formation is zero~\cite{Horodecki_fully}.

We can see that due to randomness in the initial state the evolution of entanglement between two nearest-neighbor qubits shows universal-like behavior, that is time dependence which is to a large extend independent of the precise form of the underlying nearest-neighbor hamiltonian which generates evolution. Universality is not exact, there are still ``signatures'' of specific hamiltonian at intermediate times (for instance, compare curves for the Heisenberg and Ising model in Fig.~\ref{fig:H}), overall though the dependence is similar. Bell-like shape of entanglement is a consequence of two competing processes. On the one hand, non-separable evolution naturally tends to produce entanglement from an initially separable state while on the other hand, it will tend to destroy it because $\ket{\psi(t)}$ will approach a random state as time grows and the reduced density matrix will approach an identity matrix having zero entanglement~\cite{iden}. The later process of destroying entanglement has two sources: first, randomness of the initial state is spread out throughout the system and second, dynamics itself will tend to produce random state. Universality is a consequence of randomness in the initial state, \ie, of its generic separable form. For each hamiltonian there are rare specific separable initial states for which deviations from the above average behavior will be large.

As we have seen in Fig.~\ref{fig:H} two-body random matrix model, which is a parameter-free model, describes evolution of entanglement sufficiently well also for other systems. Therefore, from now on we are going to focus on a two-body random matrix model, studying more precisely how the entanglement between two qubits evolves with time.  

\subsection{Two-body random matrix model}
\begin{figure}[h]
\centerline{\includegraphics[width=3.3in]{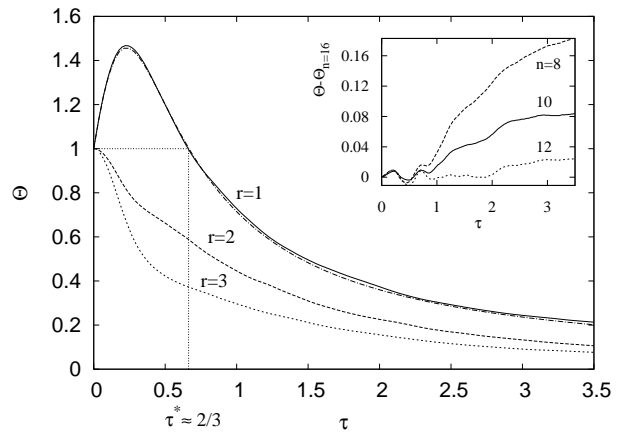}}
\caption{Average $\Theta(\tau)$ (\ref{eq:taudef}) for nearest-neighbor qubits ($r=1$), next nearest ($r=2$) and qubits separated by 2 qubits ($r=3$). For directly coupled qubits, $r=1$, states on average can not be used for distillation for times larger than $\tau^* \approx 2/3$. Chain curve (almost overlapping with the full curve for $r=1$) is $\Theta=\frac{1+4\tau}{1+6\tau^2}$. All is for two-body random matrix model with $n=16$ qubits and initial product states (\ref{eq:prodic}). In the inset we show results for $r=1$ and smaller systems.}
\label{fig:f}
\end{figure}
In Fig.~\ref{fig:f} we again show dependence of $\Theta$ (\ref{eq:taudef}) on time for a two-body random matrix model. In addition to entanglement between nearest-neighbor qubits (denoted by distance $r=1$) we also show entanglement for next-nearest neighbors ($r=2$) and qubits separated by two other qubits ($r=3$). As one can see, for qubits which are not directly coupled by the hamiltonian, \ie, $r=2$ and $r=3$ cases, $\Theta$ is always less than $1$. This happens because the production of entanglement depends on higher order terms, \eg, for $r=2$ terms of form $h_{i,i+1} h_{i+1,i+2}$ are needed, whereas for nearest-neighbors a single term $h_{i,i+1}$ is already sufficient to entangle two qubits. As a consequence, entanglement production is slower the larger is the distance between qubits while entanglement destruction due to randomness is approximately independent of the distance. In all cases we show data for $n=16$ for which finite size effects are already small. For instance, the difference between $\tau^*$ for $n=16$ and $n=18$ is $0.01$ in the case of $r=1$. In the figure we also plot rational function $\Theta(\tau)=\frac{1+4\tau}{1+6\tau^2}$ which almost perfectly overlaps with the numerics for $r=1$. Note that for short times this of course agrees with our perturbative result (\ref{eq:CLR}), seen as the initial line with slope $4$ in Figs.~\ref{fig:H} and~\ref{fig:f}. One would be tempted to think that the dependence beyond this short time, therefore also the above rational function, could be explained by higher order perturbation theory. Unfortunately it is not so. This rational dependence can not be explained by higher order perturbative calculation. In fact, we do not have any theoretical explanation for this almost perfect fit. Although going to perturbations of 2nd order in time will result in a rational function with the denominator and numerator being polynomials of order $2$ in $\tau$, the coefficients of polynomials are wrong. We have numerically checked that next perturbative orders also do not improve the situation. It therefore seems that the functional dependence of $\Theta$ (as well as of $C$ and $\lambda_{\rm min}^{T_A}$) for two-body random matrix model is essentially non-perturbative. This is in contrast with, for instance, purity or fidelity decay of initial pure states in the presence of weak coupling where perturbative approaches have been very successful~\cite{echo,Pineda:07}.       
\begin{figure}[h!]
\centerline{\includegraphics[width=3.3in]{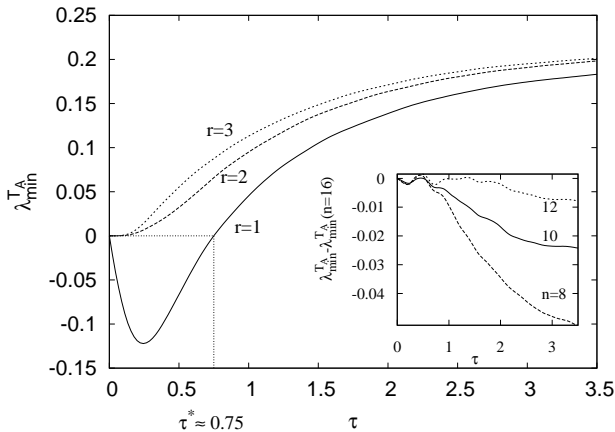}}
\caption{Average minimal eigenvalue $\lambda_{\rm min}^{T_A}$ of partially transposed density operator. We show results for nearest-neighbor qubits ($r=1$), and those for qubits separated by one ($r=2$) and two ($r=3$) qubits. All is for two-body random matrix model with $n=16$ qubits and initial product states (\ref{eq:prodic}). In the inset we show results for $r=1$ and smaller systems.}
\label{fig:lmin}
\end{figure}

\begin{figure}[h!]
\centerline{\includegraphics[width=3.3in]{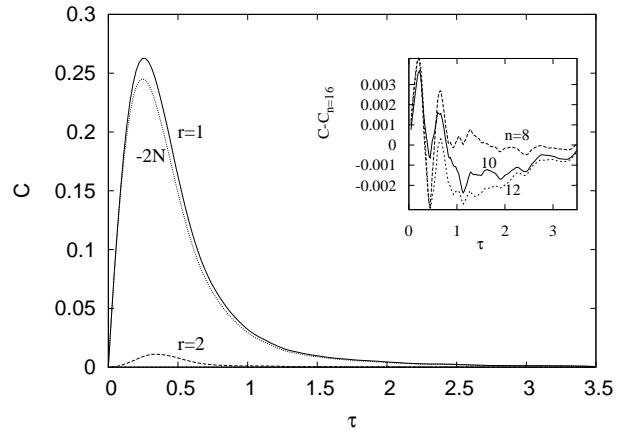}}
\caption{Average concurrence (\ref{eq:Cdef}) for nearest-neighbor qubits ($r=1$) and next nearest-neighbors ($r=2$). With dotted line we also show negativity, $-2\cdot N(\rho)$ (\ref{eq:negdef}), displaying essentially the same behavior as concurrence. For two-body random matrix model with $n=16$ qubits and initial product states.}
\label{fig:c}
\end{figure}

In Fig.~\ref{fig:lmin} we show numerical results for the average $\lambda_{\rm min}^{T_A}$. Overall, the dependence is very similar as for $\Theta(\tau)$, the only difference being that the time when $\lambda_{\rm min}^{T_A}$ becomes positive and the state gets separable is $\tau^* \approx 0.75$. In Fig.~\ref{fig:c} similar results are shown for concurrence and negativity. We can see that negativity and concurrence are almost the same for nearest-neighbor qubits. For next-nearest neighbor qubits ($r=2$) concurrence is this time non-zero but small as opposed to $\lambda_{\rm min}^{T_A}$ which is always positive. For qubits further apart, for instance $r=3$, concurrence is below the level of statistical fluctuations. 

We have checked that similar results are obtained also for other topologies of the coupling between qubits, \ie, other than nearest-neighbor. In all situations entanglement is present for two qubit reduced density matrices between qubits directly coupled by the hamiltonian, whereas entanglement is small or zero for qubits which are not directly coupled. In all cases entanglement on average disappears after finite time. Difference from sudden death of entanglement phenomenon~\cite{ESD} is that our two qubits are coupled and start from an initially separable state. In addition, our system is conservative, that is we have a finite ``environment''. 

\begin{figure}[h]
\centerline{\includegraphics[width=3.3in]{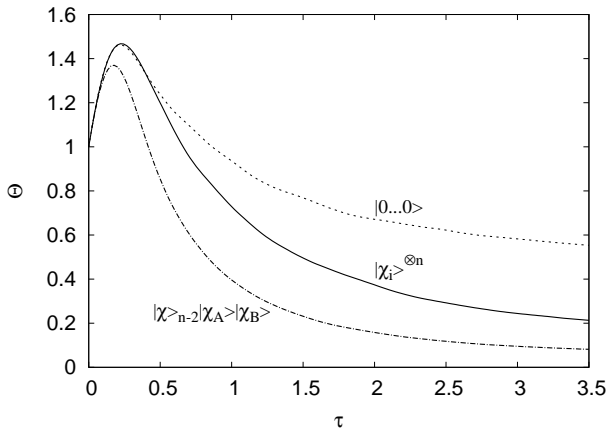}}
\caption{Average $\Theta(\tau)$ for various choices of initial states. Full line is for initial product states (\ref{eq:prodic}) (same data as in Fig.~\ref{fig:f}), chain curve is for initial state which is of product form just on the two qubits involved (\ref{eq:prodr}), while dotted curve is for homogeneous initial state $\ket{0\ldots 0}$. All is for two-body random matrix model with $n=16$ qubits. Times when average $\Theta(\tau)<1$ are $0.42, 0.66$ and $0.86$.}
\label{fig:fo}
\end{figure}

So far we have always used product initial states, where states $\ket{\chi_i}$ on individual qubits were independent. As a final numerical calculation let us check how the results depend on the choice of an initial state. Because results are similar for all quantities studied we are going to show only $\Theta(\tau)$. Besides product initial state (\ref{eq:prodic}) we used product initial state with states $\ket{\chi_i}$ on all qubits being the same. Because two-body random matrix model is invariant for single qubit rotations this is equivalent as choosing state $\ket{0\ldots 0}$ for the initial state and averaging over an ensemble of two-body random matrices. Second choice is an initial state which is of product form on the two qubits used for entanglement calculation and random on the remaining $n-2$ qubits,
\begin{equation}
\ket{\psi(0)}=\ket{\chi}_{n-2} \otimes \ket{\chi_A}\otimes \ket{\chi_B},
\label{eq:prodr}
\end{equation}
where $\ket{\chi_{A,B}}$ are random single qubit states and $\ket{\chi}_{n-2}$ is a random state of $n-2$ qubits. Numerical results are shown in Fig.~\ref{fig:fo}. We can see that, expectedly, entanglement decays slower for homogeneous initial state, $\ket{0\ldots 0}$, while it decays faster for the initial state having a full (nonseparable) random state on $n-2$ qubits (\ref{eq:prodr}). Figuratively speaking one can say that the decay of entanglement is faster the more randomness there is in the initial state.

\section{Conclusion}
We have studied how the entanglement between two qubits evolves with time when two qubits are part of a larger spin chain. Starting from an initial separable random product state entanglement first increases, reaching a maximal value, after which it decays to zero resulting in a separable state after finite time. Therefore, starting from a generic initial state possessing some randomness, two qubit reduced density matrix is on average entangled only for finite time and only for qubit pairs which are directly coupled by the hamiltonian. Time dependence of entanglement is almost independent of the specifics of the hamiltonian used for time evolution, being integrable or chaotic, and is well described by a two-body random matrix model. Results can be interpreted also in another way: it is hard to generate entanglement regardless of the dynamics if there is some randomness present in the initial state. This can be used to explain the lack of entanglement in small subsystems for generic initial conditions. The author would like to thank T.~Prosen for reading the manuscript and anonymous referee for valuable suggestions and for bringing to our attention Refs.~\cite{Sen:06} and~\cite{Dur:05}.

\end{document}